\documentclass{article} 

\newcommand{\ket}[1]{|#1\rangle}

\usepackage{amssymb}
\usepackage{amsmath}
\usepackage{natbib}
\usepackage[dvips]{graphicx}
\begin{document}

\title{Non-Contextuality, Finite Precision Measurement and the Kochen-Specker Theorem}

\author{Jonathan Barrett\\
Physique Th\'{e}orique CP 225,\\ Universit\'{e} Libre
de Bruxelles,\\
Bvd. du Triomphe,\\
1050 Bruxelles, Belgium\\
and\\
Th\'{e}orie de l'Information et des Communications CP 165/59,\\
Universit\'{e} Libre de Bruxelles,\\
Av. F.~D.~Roosevelt 50,\\ 
1050 Bruxelles, Belgium\\ 
\and Adrian Kent\\
Centre for Quantum Computation,\\ 
Department of Applied Mathematics and Theoretical Physics,\\ 
University of Cambridge,\\
Wilberforce Road, Cambridge CB3 0WA\\
United Kingdom}

\date{August 2003, revised November 2003}

\maketitle

\begin{abstract}
Meyer originally raised the question of whether non-contextual hidden
variable models can, despite the Kochen-Specker theorem, simulate the
predictions of quantum mechanics to within any fixed finite experimental
precision (Meyer, D. 1999. Phys. Rev. Lett., 83, 3751-3754). Meyer's result
was extended by Kent (Kent, A. 1999. Phys. Rev. Lett., 83, 3755-3757).
Clifton and Kent later presented constructions of non-contextual hidden
variable theories which, they argued, indeed simulate quantum mechanics in
this way (Clifton, R and Kent, A. 2000. Proc. Roy. Soc. Lond. A, 456,
2101-2114).

These arguments have evoked some controversy.
Among other things, it has been suggested that the Clifton-Kent models do 
not in fact reproduce correctly the predictions of quantum mechanics,
even when finite precision is taken into account.  It has also 
been suggested that careful analysis of the notion
of contextuality in the context of finite precision measurement motivates 
definitions which imply that the Clifton-Kent models are in fact 
contextual.  Several critics have also argued that the
issue can be definitively resolved by experimental tests of the Kochen-Specker
theorem or experimental demonstrations of the contextuality of Nature.  

One aim of this paper is to respond to and rebut criticisms of the
Meyer-Clifton-Kent papers.  We thus elaborate in a little more detail how the Clifton-Kent models
can reproduce the predictions of quantum mechanics to arbitrary
precision.  We analyse in more detail the relationship between
classicality, finite precision measurement and contextuality, 
and defend the claims that the Clifton-Kent models are both essentially classical and 
non-contextual.  We also examine in more detail the senses in which a 
theory can be said to be contextual or non-contextual, and in which
an experiment can be said to provide evidence on the point.  
In particular, we criticise the suggestion that a decisive experimental
verification of contextuality is possible, arguing that the idea rests
on a conceptual confusion.   


\end{abstract}

\section{Introduction}\label{introduction}

\subsection{The Kochen-Specker theorem}

Consider a set ${\cal K}$ of Hermitian operators that act on an
$n$-dimensional Hilbert space. Suppose that $V$ is a map that takes a
Hermitian operator in ${\cal K}$ to a real number in its spectrum. We call
such a map a \emph{colouring} of ${\cal K}$. If the following conditions are
satisfied
\begin{eqnarray}\label{kscriteria}
V(\hat{A}+\hat{B})&=&V(\hat{A})+V(\hat{B})\nonumber\\
V(\hat{A}\hat{B})&=&V(\hat{A})V(\hat{B})\nonumber\\
&&\ \ \ \ \forall \hat{A},\hat{B}\in {\cal K} \ \mathrm{such\ that\ }
[\hat{A},\hat{B}]=0,
\end{eqnarray}
then the map is a \emph{KS-colouring} of ${\cal K}$.  We call these
conditions the KS criteria. Kochen and Specker's celebrated
theorem \citep{specker,ks} states that if $n>2$ there are
\emph{KS-uncolourable} sets, i.e., sets ${\cal K}$ for which no KS-colouring
exists. It follows trivially that the set of all Hermitian operators acting
on a Hilbert space of dimension $>2$ is KS-uncolourable.

The fact that the set of all Hermitian operators in dimension $>2$ is
KS-uncolourable is a corollary of Gleason's theorem \citep{gleason}. This was
first pointed out in \citet{bell}, where an independent proof was also given.
Kochen and Specker constructed the first finite KS-uncolourable set. Many
proofs along the lines of Kochen and Specker's have since been produced by
constructing demonstrably KS-uncolourable sets \citep[see, e.g.,
][]{peresbook,zimbapenrose,conwaykochen,ceg}.  The most common type of proof
describes a set of $1$-dimensional projection operators in $n$ dimensions that
is KS-uncolourable. If we represent $1$-dimensional projections by vectors
onto which they project, and colour the corresponding set of vectors with a
$1$ or a $0$, the KS criteria would imply that for each orthogonal $n$-tuple
of vectors, exactly one must be coloured $1$, and all the rest $0$. The
Kochen-Specker theorem can then be proved by showing that the colouring
condition cannot be satisfied. In their original proof, Kochen and Specker
describe a set of $117$ vectors in $3$ dimensions that is
KS-uncolourable.\footnote{How small can a KS-uncolourable set of vectors be?
  The current records stand at $31$ vectors in $3$ dimensions
  \citep{conwaykochen} and $18$ in $4$ dimensions \citep{ceg}.}

Of course, the well-known proofs of the Kochen-Specker theorem referred to
above are logically correct. Moreover, the Kochen-Specker theorem undeniably
says something very important and interesting about fundamental physics: it
shows that the predictions of quantum theory for the outcomes of measurements
of Hermitian operators belonging to a KS-uncolourable set cannot be precisely
reproduced by any hidden variable theory that assigns real values to these
operators in a way that respects the KS criteria, since no such hidden
variable theory exists.  However, debate continues over the extent to which
Kochen and Specker succeeded in their stated aim, ``to give a proof of the
nonexistence of hidden variables''\citep[p.59]{ks}, even when this is
qualified (as it must be) by restricting attention to non-contextual hidden
variables.  Before summarising and continuing this debate, we review why one
might be interested in hidden variable theories in the first place.

Consider a system in a state $\ket{\psi}$ and a set of observables $A, B,
C,\ldots$ such that $\ket{\psi}$ is not an eigenstate of $\hat{A}, \hat{B},
\hat{C},\ldots$; here we use capital letters with hats to denote Hermitian
operators and capital letters without hats to denote the corresponding
observables. Orthodox quantum mechanics leads us to say something like this:
if we measure $A$, we will obtain the result $a$ with probability $p_a$, if
we measure $B$, we will obtain the result $b$ with probability $p_b$, and so
on. With an ease born of familiarity, the well trained quantum mechanic will
not bat an eyelid at such statements. But, one might well ask: why are they
so oddly phrased? Could this just be a rather awkward way of saying that with
probability $p_a$, the value of $A$ \emph{is} $a$, or with probability $p_b$,
the value of $B$ \emph{is} $b$, and so on?

Suppose that the set $A, B, C,\ldots$ corresponds to a KS-uncolourable set of
operators $\hat{A}, \hat{B}, \hat{C},\ldots$.  The suggestion is that at a
given time, each observable in the set has some definite value associated
with it, defined by some ``hidden'' variables of the system. The significance
of the KS criteria is that if the Hermitian operators associated with two
observables commute, then according to quantum mechanics, the observables can
be simultaneously measured, and the values obtained will satisfy the KS
criteria (and in general will satisfy any functional relationships that hold
between the operators themselves).  We are not logically compelled to assume
that any hidden variable theory shares these properties. However, the
standard motivation for considering hidden variables is to examine the
possibility that quantum theory, while not incorrect, is incomplete, Thus
motivated, it seems natural to assume that the colouring defined by the
hidden variables must also satisfy the KS criteria. But given this
assumption, since there is no such colouring, the original supposition that
the observables have definite values must be wrong.

The contradiction obtained in the Kochen-Specker theorem is avoided if,
instead of defining a map $V$, we assign values to Hermitian operators in
such a way that the value assigned to a particular Hermitian operator depends
on which commuting set we are considering that operator to be part of. Such a
value assignment is called \emph{contextual}. Hidden variable interpretations
of quantum theory based on contextual value assignments can be defined. In
such {contextual hidden variable} (CHV) interpretations, the outcome
obtained on measuring a certain quantum mechanical observable is indeed
pre-defined, but depends in general on which other quantum mechanical
observables are measured at the same time. Thus, if we take the KS criteria
for granted, Kochen and Specker's results show that there are no {\it
non-contextual hidden variable} (NCHV) interpretations of the standard
quantum mechanical formalism.

It may seem tempting to phrase this more directly, concluding that the
Kochen- Specker theorem shows that Nature cannot be described by any
non-contextual hidden variable theory.  Another possible conclusion is that
the KS theorem implies that we could exclude non-contextual hidden variable
theories if the predictions of quantum theory were confirmed in a suitably
designed experiment.  We will argue below that neither conclusion is correct.

\subsection{Querying the scope of the KS theorem}

We next review some earlier discussions that suggest limitations on what can
be inferred from the Kochen-Specker theorem.

Some time ago, Pitowsky devised models \citep{pitowskyprd,pitowskyphilsci}
that assign values non-contextually to the orthogonal projections in three
dimensions and nonetheless satisfy (\ref{kscriteria}) ``almost everywhere''
\citep[p.2317]{pitowskyprd}.  The models are non-constructive, requiring the
axiom of choice and the continuum hypothesis (or some suitable weaker
assumption) for their definition. Another complication is that the term
``almost everywhere'' is not meant in the standard sense, but with respect to
a non-standard version of measure theory proposed by Pitowsky
\citep[see][]{pitowskyprd} which, among other disconcerting features, allows
the intersection of two sets of probability measure $1$ to have probability
measure $0$ \citep{pitowskyprl}.

Pitowsky's models disagree with quantum mechanics for some measurement
choices, as the KS theorem shows they must. They thus do not {\it per
  se} seem to pose an insuperable obstacle to arguments that --- either
directly from the theorem or with the aid of suitable experiments --- purport
to demonstrate the contextuality of Nature.\footnote{Nor, it should be
stressed, did Pitowsky suggest that
  they do.}  After all, either the demonstration of a finite
non-colourable set of projectors is sufficient to run an argument, or it is
not. If it is, Pitowsky's models are irrelevant to the point; if it is not,
it is not obvious that the models, equipped as they are with an entirely
novel version of probability theory, are either necessary or sufficient 
for a refutation. 

A more direct challenge to the possibility of theoretical or experimental
refutations of non-contextual hidden variables was presented in
\citet{meyer}, where the implications of finite experimental
precision are emphasised: ``Only finite precision measurements are experimentally
reasonable, and they cannot distinguish a dense subset from its
closure'' \citep[p.3751]{meyer}. Meyer identified a particularly simple and elegant
construction, originally described in \citet{godsilzaks}, of a
KS-colourable dense subset of the set of projectors in three
dimensions.\footnote{
  As Pitowsky has since noted, Meyer's argument could also be framed using
  one of Pitowsky's
  constructions of dense KS-colourable sets of projectors rather than
  Godsil and Zaks'.}  His conclusion was that, at least in three
dimensions, the Kochen-Specker theorem could be ``nullified''.\footnote{It
  should perhaps be emphasised that the sense
  of ``nullify'' intended here is ``counteract the force or
  effectiveness of'', not ``invalidate''. Neither Meyer nor anyone
  else has suggested that the proofs of the Kochen-Specker theorem are
  flawed.} As a corollary, Meyer argued that the KS theorem alone
cannot discriminate between quantum and classical (therefore non-contextual)
information processing systems.  

Meyer left open the question of whether static non-contextual hidden variable
theories reproducing the predictions of quantum theory for three dimensional
systems actually exist: his point was that, contrary to most previous
expectations, the Kochen-Specker theorem does not preclude such hidden
variable theories.

Meyer's result was subsequently extended by a construction of KS-colourable
dense sets of projectors in complex Hilbert spaces of arbitrary dimension
\citep{akprl}. Clifton and Kent (CK) extended the result further by
demonstrating the existence of dense sets of projection operators, in complex
Hilbert spaces of arbitrary dimension, with the property that no two
compatible projectors are members of incompatible resolutions of the
identity \citep{ck}. The significance of this property is that it makes it trivial to
construct a distribution over different hidden states that recovers the
quantum mechanical expectation values.  Such a distribution is, of course,
necessary for a static hidden variable theory to reproduce the predictions of
quantum theory.  Similar constructions of dense subsets of the sets of all
positive operators were also demonstrated \citep{akprl,ck}.  CK
presented their constructions as non-contextual hidden variable
theories that can indeed simulate the predictions of quantum mechanics in
the sense that the theories are indistinguishable in real experiments in
which the measurement operators are defined with finite precision.

The arguments set out by Meyer, Kent, and Clifton and Kent (MKC) have evoked
some 
controversy \citep[see, e.g.,
][]{mermin99,cabello,basuetal,sbz,larsson,peres,appleby00,appleby01,appleby02,appleby03,havliceketal,cabellopra,breuer} and even a parody\citet{peresparody}.
Among other things, it has been suggested \citep{cabellopra} that the CK
models do not in fact reproduce correctly the predictions of quantum
mechanics, even when finite precision is taken into account. It has also been
suggested \citep{sbz,larsson,appleby00} that careful analysis of the notion of
contextuality in the context of finite precision measurement motivates
definitions which imply that the CK models are in fact contextual.

Several of these critiques raise novel and interesting points, which have
advanced our understanding of the Kochen-Specker theorem and its implications.  
Nonetheless, we remain convinced that the essential insight of \citet{meyer}
and all the substantial points made in \citet{akprl} and \citet{ck} are valid.
One aim of this paper is thus to respond to and rebut MKC's critics. 

Perhaps unsurprisingly, quite a few critics have made similar points.  
Also, some purportedly critical arguments make points irrelevant to 
the arguments of the MKC papers (which were carefully limited in
their scope).  Rather than producing a comprehensive --- but, we fear,
unreadable --- collection of counter-critiques of each critical article, 
we have tried in this paper to summarise and comment on the most 
interesting new lines of argument.  

Among other things, we explain here in a little more detail how the CK models can
reproduce the predictions of quantum mechanics to arbitrary precision, both
for single measurements and for sequences.  We point out a conceptual
confusion among critics who suggest that the models are contextual, noting
that the arguments used would (incorrectly) suggest that Newtonian physics
and other classical theories are contextual. We also defend the claim that
the CK models are essentially classical. Indeed, as we explain, the models
show in principle that one can construct classical devices that assign
measurement outcomes non-contextually and yet simulate quantum mechanics to
any given fixed non-zero precision.  In
summary, we reiterate the original claim of MKC that the models, via finite
precision, provide a loophole --- which is physically implausible but
logically possible --- in the Kochen-Specker argument.

Running through these debates is another theme: the alleged possibility of
experimental tests of the Kochen-Specker theorem, or experimental
demonstrations of the contextuality of Nature.  
Quite a few experiments purporting to test contextuality have recently 
been proposed \citep{cg,szwz,basuetal} and performed 
\citep{hlzpg,hlbbr}.  Several authors have suggested an analogy between 
these purported experimental tests of contextuality and Bell experiments
testing local causality.  

Another aim of this paper is to go beyond previous
discussions in examining in detail the senses in which a theory can be said to
be contextual or non-contextual, and in which an experiment can be said to
provide evidence for these. Broadly, we are critical of the idea of an
experimental test of non-contextuality, arguing that the idea rests on
conceptual confusion. The experiments that have been performed
test predictions of quantum mechanics which certainly conflict with
some classical intuitions, and which might indeed raise questions
about the contextuality of measurements to someone familiar 
only with certain aspects of quantum theory.    
But, as we re-emphasize in this paper, they certainly do not provide 
loophole-free demonstrations of the contextuality of Nature, since
the CK models can reproduce the experimental data.  

There is also a more basic problem.
Interpreting the experiments  
in a way which raises the question of contextuality at all  
requires assuming significant parts of the formalism of quantum theory.
On the other hand, if we simply assume quantum theory is
valid, without any qualification, we need no experiment: the Kochen-Specker theorem already 
excludes non-contextual hidden variable theories.  
It is thus quite hard to pin down what exactly a purported
experimental test of contextuality proves, or could prove, that 
we do not know already.  In our opinion, this key point is not adequately 
addressed in the papers under discussion \citep{cg,szwz,basuetal,hlzpg,hlbbr}.  

\section{MKC models}\label{MKCmodels}

Kochen and Specker's declared motivation for constructing finite
uncolourable sets is interesting, both because it partly anticipates the point
made a third of a century later by Meyer and because its implications seem to
have been largely ignored in the period intervening:

\begin{quote}
  It seems to us important in the demonstration of the non-existence
  of hidden variables that we deal with a small finite partial Boolean
  algebra. For otherwise a reasonable objection can be raised that in
  fact it is not physically meaningful to assume that there are a
  continuum number of quantum mechanical propositions. \citep[p.70]{ks}
\end{quote}

What Kochen and Specker neglected to consider is that the objection might be
sharpened: it could be that in fact only a specified countable set of quantum
mechanical propositions exist, and it could be that this set has no
KS-uncolourable subsets (finite or otherwise). This is the possibility that
the MKC models exploit.

Before discussing these models, we wish to reemphasise the disclaimers made
in \citet{ck}. MKC models describe a type of hidden variable theory that
is a logically possible alternative to standard quantum theory, but not, in
our view, a very plausible one. The CK constructions in particular, are ugly
and contrived models, produced merely to make a logical point.  One might
hope to devise prettier hidden variable models which do the same job, using a
colouring scheme as natural and elegant as Godsil and Zaks'.  Even if such models
were devised, though, we would not be inclined to take them too seriously as
scientific theories.

However, we think it important to distinguish between scientific
implausibility and logical impossibility. The models show that only the
former prevents us from adopting a non-contextual interpretation of any real
physical experiment. Another reason for studying the models --- in fact,
Meyer's main original motivation \citep{meyer} --- is to glean insights into the
possible r\^ole of contextuality in quantum information theory.

\subsection{Projective measurements}

The argument of \citet{ks}, and most later discussions until
recently, including \citet{meyer}, assume that the quantum theory of measurement can be
framed entirely in terms of projective measurements. This remains a tenable
view, so long as one is willing to accept that the experimental configuration
defines the quantum system being measured.\footnote{For instance, a
projective measurement on a quantum system $S$ together with an ancillary
  quantum system $A$ requires us, on this view, to take $S+A$ as the
  system being measured, rather than speaking of a positive operator valued measurement
  being carried out on $S$.}  We adopt it here, postponing discussion
of positive operator valued (POV) measurements to the next subsection.

Meyer identified a KS-colouring, originally described in \citet{godsilzaks},
of the set $S^2\cap Q^3$ of unit vectors in $R^3$ with rational components, or
equivalently of the projectors onto these vectors. As he pointed out, not only
is this set of projectors dense in the set of all projectors in $R^3$, but the
corresponding set of projective decompositions of the identity is dense in the
space of all projective decompositions of the identity.

Meyer's result is enough to show that an NCHV theory along these lines is not
ruled out by the Kochen-Specker theorem. It does not show that such a theory
exists. For this we need there to be KS-colourable dense sets of projectors
in complex Hilbert spaces of arbitrary dimension. Further, it is not enough
for each set to admit at least one KS-colouring. For each quantum state, one
must be able to define a distribution over different KS-colourings such that
the correct quantum expectation values are obtained. For these reasons, Kent
extended Meyer's result by constructing KS-colourable dense sets of
projectors in complex Hilbert spaces of arbitrary dimension \citep{akprl}.
Clifton and Kent extended the result further \citep{ck} by demonstrating the
existence of dense sets of projection operators, in complex Hilbert spaces of
arbitrary dimension, with the property that no two compatible projectors are
members of incompatible resolutions of the identity. The significance of this
property is that it makes it trivial to construct a distribution over
different hidden states that recovers the quantum mechanical expectation
values.

CK argue that this construction allows us to define a non-contextual
hidden variable theory that simulates quantum mechanics, by the following
reasoning.  First, let us suppose that, as in the standard von Neumann
formulation of quantum mechanics, every measurement corresponds to a
projective decomposition of the identity. However, because any experimental
specification of a measurement has finite precision, we need {\it not}
suppose that every projective decomposition corresponds to a possible
measurement.  Having defined a dense set of projectors ${\cal P}$ that gives
rise to a dense set of projective decompositions of the identity ${\cal D}$,
we may stipulate that every possible measurement corresponds to a
decomposition of the identity in ${\cal D}$.  The result of any measurement
is determined by hidden variables that assign a definite value to each
operator in ${\cal P}$ in a non-contextual manner. Via the spectral
decomposition theorem, those Hermitian operators whose eigenvectors
correspond to projectors in ${\cal P}$ are also assigned values. If
measurements could be specified with infinite precision, then it would be
easy to distinguish this alternative theory from standard quantum mechanics.
We could simply ensure that our measurements correspond exactly to the
projectors featured in some KS-uncolourable set. If they in fact corresponded
to slightly different projectors, we would detect the difference.

Now, for any finite precision, and any KS-uncolourable set of projectors,
there will be projectors from ${\cal P}$ sufficiently close that the
supposition that our measurements correspond to those from ${\cal P}$ will
not make a detectable difference. So, which particular element of ${\cal D}$
does this measurement correspond to? CK propose that the answer to this
question is determined algorithmically by the hidden variable theory.

Let us illustrate how this could work by fleshing out, with more detail
than was given in \citet{ck}, one way in which a CK model could work.  
Consider some ordering $\{ d^1 , d^2 ,
\ldots \}$ of the countable set ${\cal D}$.  Let $\epsilon$ be a parameter
much smaller than the precision attainable in any current or foreseeable
experiment.  More precisely, $\epsilon$ is sufficiently small that it will be
impossible to tell from the outcome statistics if a measurement attempts to
measure a decomposition $d = \{ P_1 , \ldots , P_n \}$ and actually measures
a decomposition $d' = \{ P'_1 , \ldots , P'_n \}$, provided $ | P_i - P'_i |
< \epsilon$ for all $i$. Suppose now we design a quantum experiment which
would, if quantum theory were precisely correct, measure the projective
decomposition $d$. (Of course, the experimenter can only identify $d$ to within the limits
of experimental precision, but, on the hypothesis that all measurements are
fundamentally projective, we suppose that in reality the value of $d$ is an
objective fact.)  We could imagine that the hidden variable theory uses the
following algorithm: first, it identifies the first decomposition $d^i =\{
P^i_1 , \ldots , P^i_n \} $ in the sequence such that $ | P_j - P^i_j | <
\epsilon$ for all $j$ from $1$ to $n$. Then, it reports the outcome of the
experiment as that defined by the hidden variables for $d^i$: in other words,
it reports outcome $j$ if the hidden variable theory ascribes value $1$ to
$P^i_j$ (and hence $0$ to the other projectors in $d$).

It may be helpful to visualise this sort of model applied to projectors in
three real dimensions.  The system to be measured can be pictured as a sphere
with (infinitesimally thin) spines of some fixed length sticking out along
all the vectors corresponding to projectors in $D$, coloured with $1$ or $0$
at their endpoint.  A quantum measurement defines an orthogonal triple of
vectors, which in general is not aligned with an orthogonal triple of spines.
Applying the measurement causes the sphere to rotate slightly, so that a
nearby orthogonal triple of spines becomes aligned with the measurement
vectors.  The measurement outcome is then defined by the spine colourings.

Some points are worth emphasising here.  First, the algorithm we have just
described obviously {\it cannot} be obtained from standard quantum theory. It
is the hidden variable theory that decides which projective decomposition is
actually measured.  Some critics have implicitly (or explicitly) assumed that
the measured decomposition must be precisely identified by standard quantum
theoretic calculations.\footnote{See, for example, \citet{peres}, where the
``challenge'' seems to be based on a misunderstanding of this point
and on neglect of the POV models defined in the next section, and also
\citet{appleby01}.}  But finite precision hidden variable
models need not be so constrained: all they need to do is simulate quantum
theory to within finite precision.

Second, as the algorithm above suggests, any given CK model actually contains
an infinite collection of sub-models defined by finite subsets $\{ d^1 , d^2
, \ldots , d^r \}$ of $D$ with the property that they are able to reproduce
quantum theory to within some finite precision $\epsilon_r$, where $\epsilon_r
\rightarrow 0$ as $r \rightarrow \infty$.  At any given
point in time, there is a lower bound on the precision actually attainable in
any feasible experiment. Hence, at any given point in time, one (in fact
infinitely many) of the finite sub-models suffices to reproduce quantum
theory to within attainable experimental precision.  In other words, at any
given point in time, MKC's argument can be run without using infinite dense
subsets of the sets of projectors and projective decompositions.

Third, we recall that the models CK originally defined are not complete
hidden variable models, since no dynamics was defined for the hidden
variables.  As CK noted, the models can be extended to cover sequential
measurements simply by assuming that the hidden variables undergo a
discontinuous change after a measurement, so that the probability
distribution of the post-measurement hidden variables corresponds to that
defined by the post-measurement quantum mechanical state vectors.  A complete
dynamical non-contextual hidden variable theory needs to describe successive
measurements in which the intervening evolution of the quantum state is
non-trivial.  In fact (though CK did not note it), this could easily be done,
by working in the Heisenberg rather than the Schr\"odinger picture, and
applying the CK rules to measurements of Heisenberg operators.
In this version of the CK model, the hidden
variables define outcomes for measurements, change discontinuously so
as to reproduce the probability distributions for the transformed quantum
state, and then remain constant until the next measurement.

\subsection{Positive operator valued measurements}

Dealing with projective measurements is arguably not enough.  One quite
popular view of quantum theory holds that a correct version of the
measurement rules would take POV measurements as fundamental, with projective
measurements either as special cases or as idealisations which are never
precisely realised in practice.  In order to define an NCHV theory catering
for this line of thought, Kent constructed a KS-colourable dense
set of positive operators in a complex Hilbert space of arbitrary dimension,
with the feature that it gives rise to a dense set of POV decompositions of
the identity \citep{akprl}. Clifton and Kent constructed a dense set of positive
operators in complex Hilbert space of arbitrary dimension with the special
feature that no positive operator in the set belongs to more than one
decomposition of the identity \citep{ck}. Again, the resulting set of POV decompositions
is dense, and the special feature ensures that one can average over hidden
states to recover quantum predictions.  Each of the three points made at the
end of the last section applies equally well to the POV models.

We should stress that the projective and POV hidden variable models defined
in \citet{akprl} and \citet{ck} are separate theories. One can consider whichever
model one prefers, depending whether one is most interested in simulating
projective or POV quantum measurements, but they are not meant to be
combined.  The POV hidden variable model does, as of course it must, define
outcomes for projective measurements considered as particular cases of POV
measurements --- but not in the same way that the projective hidden variable
model does.

The CK models for POV measurements have, surprisingly, been neglected by some
critics \citep[e.g.,][]{peres}, who object to the CK projective models on the
grounds that they unrealistically describe outcomes of ideal but imprecisely
specified projective measurements.  As we noted above, this objection is indeed
reasonable if one takes the view that one should define the measured quantum
system in advance, independent of the details of the measurement apparatus,
or if one regards POV measurements as fundamental for any other reason. The
POV measurement models were devised precisely to cover these points.

\section{Some criticisms of the MKC models}\label{furtherdiscussion}

\subsection{Are the CK models classical?}\label{classicality}

Clifton and Kent claimed that the CK models show ``there is no truly
compelling argument establishing that non-relativistic quantum mechanics
describes classically inexplicable physics'' \citep[p.2113]{ck}.  Some
\citep{appleby00,appleby01,appleby03,havliceketal} have queried whether the
models can, in fact, properly be described as classical, given that they
define values on dense subsets of the set of measurements in such a way that
every neighbourhood contains operators with both truth values. This feature
implies that the models do not satisfy what we call the \emph{faithful
  measurement condition}: that one can in general ascribe a value to an 
operator $P$, such that this value, or one close to it,
is obtained with high probability when a high precision measurement of $P$ is
performed.  

Appleby
\citep[see][]{appleby00,appleby01,appleby03}
has discussed the faithful measurement condition at some length, 
arguing that it is a necessary property of measurements
in classical models.  Appleby notes that a classical measurement tells us
some definite fact about the system as it was before measurement,
and goes on to argue that the dense --- in Appleby's words, ``radical''
or ``pathological'' \citep{appleby01,appleby03} --- discontinuities of 
truth values in the CK models mean that they cannot satisfy this epistemological 
criterion: let us call it the {\it definite revelation criterion}.      

Before considering Appleby's argument, one might first ask whether
dense discontinuities are actually necessarily a feature of any CK-type model
that simulates quantum mechanics.  As Appleby \citep{appleby01,appleby03} 
and Cabello \citep{cabellopra} show, they are.\footnote{In \citet{cabellopra}, 
 it is ostensibly argued that any model
  of the type constructed by CK must lead to experimental predictions that
  differ from those of quantum mechanics. This is clearly not correct.
  However, an examination of Cabello's argument reveals a technical assumption
  that is not true of the CK models, as noticed by Clifton in a private
  communication to Cabello, reported by Cabello in a footnote. Cabello's reply
  to Clifton is essentially an attempt to justify the assumption by appeal to
  something like the faithful measurement condition. Thus his argument is best
  viewed as a demonstration that CK-type models cannot have this feature.
  \citet{appleby01} offers a similar analysis of Cabello's argument.} 
Appleby's argument thus cannot be sidestepped.  

However, in our opinion, while the CK models clearly do not satisfy the
faithful measurement criterion, they {\it do} satisfy the definite 
revelation criterion, in the same sense that standard models in
classical mechanics do.  The CK models can thus
indeed properly be described as classical.

We believe this claim is ultimately justified 
by virtue of the phase space structure and the logical structure of
the CK models, both of which are classical.   However, since
discussion has focussed on the discontinuity of the CK models,
it is worth considering this point in more detail.     

Note, first, that discontinuity {\it per se} is clearly not an obstacle to
classicality, according to standard definitions.  
Point particles and finite extended objects with
boundary discontinuities are routinely studied in classical physics. 
Moreover, if the mere existence of discontinuities in the truth values
assigned to operators were the crucial issue, the KS
theorem would be redundant --- it is immediately obvious that any truth values
assigned by hidden variables must be discontinuous, since the only possible
truth values are $0$ and $1$, and both must be realised.
Any argument against the classicality of the CK models must, then, stem from the fact that
their discontinuities are dense. 

One possible argument against the classicality of models with 
dense discontinuities might be that, if the faithful
measurement condition is not satisfied, then little sense can be given to the
notion of one finite precision measurement being more ``precise'' than another. 
If one is not able to compare degrees of precision, it might be argued, 
one has not recovered the classical concept of measurement at all.  
In reply, we note that there is in fact a clear definition of the precision
of measurement devices within CK models.  For example, if a high precision device 
is supposed to measure $z$-spin, then it will with high probability return a 
value of $+1$ whenever a measurement is performed on a particle 
prepared (by another high precision device) in the corresponding eigenstate.
The precision of the relevant devices is then calibrated by the difference  
between the actual outcome probability and $1$, which would represent 
perfect precision.   This feature of CK models seems to have 
been neglected: for instance, it is simply not true that, as Appleby 
suggests \citep[p.6]{appleby01}, in CK models, the outcome of a measurement of 
an observable $P$ ``does not 
reveal any more information $\ldots$ [about the pre-existing value of $P$]
$\ldots$ than could be obtained by tossing a coin''.  If
an unknown quantum state drawn from a known ensemble is measured, then
obtaining a valuation for the actually measured observable $P'$ 
generally {\it does} give some statistical information about the 
pre-measurement valuation of the target observable $P$, whenever
$P$ is one of the observables to which the model assigns a valuation. 
 
Another possible argument might begin from the observation that systems that are
actually studied in the context of classical mechanics generally satisfy the faithful
measurement condition, which might suggest that the condition is implied by some part
of classical intuition.   But induction based merely on familiarity is a 
dangerous exercise.  (Three-legged dogs are still canine, for example.) 
It is, admittedly, rare to consider
classical systems which have dense discontinuities, but it does not 
contradict any standard definition of classicality of which we are aware. 
Given consistent evolution laws, one can sensibly study the
behaviour of a classical system in which point particles are initially sited
at every rational vector in $R^3$, for instance.

To address Appleby's point directly, we note that according to
the CK models a measurement {\it does}, in fact, reveal the pre-existing
valuation of an observable.  
Consider again a CK model defined by the algorithm given in section 2.1.   
It is true that finite experimental precision
makes it impossible for a human experimenter to identify precisely either 
the quantum observable which any given experiment would end up measuring
if quantum theory is correct, or the CK observable which it would end up
measuring if a CK model were correct.  Nonetheless there is, according to the CK models, a 
fact of the matter about the identity of both observables.
The process runs thus: some definite quantum observable is defined by the
experimental configuration; some definite CK observable, related to the quantum 
observable by some definite algorithm, is thus also indirectly defined by
the experimental configuration; the pre-existing valuation of this
CK observable is revealed by the experiment.  
An omniscient deity viewing the whole process ``from the outside''
could verify the action of the CK model, following (for example) the 
algorithm discussed in Section 2.1, and
predict in advance precisely which CK observable will be addressed
and the valuation that will be revealed.  
In other words, the CK models {\it do} satisfy the definite 
revelation criterion, as we understand it.  
 
Finally, but importantly, we can
offer an alternative response to those unpersuaded by any of the above arguments.  
As we noted earlier, one can define CK models
that simulate quantum mechanics adequately (given any specific attainable
experimental precision) using finite collections of projections and projective
decompositions. These models are still discontinuous, but they have only finitely
many discontinuities, rather than a dense set.  They therefore satisfy the faithful measurement
condition (this being possible because any particular such model will make
different predictions from quantum mechanics once a certain precision is
exceeded). As above, one can visualise such a model, in $R^3$, as defined by a
sphere with finitely many spines projecting from it. 
In terms of its discontinuities (which are finite in number) and its
dynamics (which could be precisely defined by a sufficiently complex
force law) such an object is analogous to a 
finite set of point particles.   There is no sensible
definition of classicality that renders it (or analogues with more
degrees of freedom) non-classical.   

\subsection{Are the CK models consistent with quantum probabilities?}

In
\citet{appleby00}, it is argued that any model of a certain type must
either be contextual or violate the predictions of quantum mechanics. In
\citet{breuer}, it is argued that NCHV models of yet a different type
make different predictions from quantum mechanics. Appleby and Breuer
both make assumptions that are not true of the CK constructions.

In \citet{appleby00}, it is argued that any non-contextual model of a certain
kind makes different predictions from quantum theory. Appleby assumes that in
an imprecise measurement of observables corresponding to three projectors, the
three projectors actually measured are not exactly commuting, but are picked
out via independent probability distributions. However, this is not how CK
models work.  For example, in a CK model for projective measurements, the
projectors actually measured are always commuting (assuming that they are
measured simultaneously) - this is one of the axioms of the theory that relate
its mathematical structure to the world, i.e., it is not some kind of
miraculous coincidence. If the projectors are measured sequentially, then the
rules of the model stipulate that the hidden state changes discontinuously
after each measurement and Appleby's analysis no longer applies. Similar
remarks apply to the POV version.

In \citet{breuer} it is shown that any finite precision NCHV
model that assigns values to a dense subset of projection operators, and also
satisfies a certain extra assumption, must make different
predictions from quantum mechanics. Suppose that a spin measurement is
performed on a spin-$1$ particle and that the measurement direction desired
by the experimenter (the target direction) is $\vec{n}$. The assumption is
that the actual measurement direction is in a random direction $\vec{m}$, and
that the distribution $\omega_{\vec{n},\epsilon}(\vec{m})$ over possible
actual directions, given $\vec{n}$ and the experimental precision $\epsilon$,
satisfies
\[
\omega_{R\vec{n},\epsilon}(R\vec{m})=\omega_{\vec{n},\epsilon}(\vec{m}),
\]
for all rotations $R$. Of course, the CK models do not satisfy this
condition, and Breuer notes this. In fact, it is clear that no model that
colours only a countable set of vectors could satisfy the condition. 
To those who regard Breuer's condition as desirable on aesthetic grounds, 
we need offer no counter-argument: it was conceded from the beginning
that the CK models are unaesthetic. 

\subsection{Non-locality and quantum logic}\label{nonlocality}

Any hidden variable theory that reproduces the predictions of quantum
mechanics must be non-local, by Bell's theorem. The CK models are no
exception. Some have argued \citep{appleby02,boyleschafir}, however, that
non-locality is itself a kind of contextuality, and that any theory that is
non-local must also, therefore, be contextual. Indeed, it is relatively common
to read in the literature the claim that non-locality is a special case of
contextuality. Here, we simply wish to point out that non-locality and
contextuality are logically independent concepts. Newtonian gravity provides
an example of a theory that is non-contextual and non-local.  One can also
imagine theories that are contextual and local - for example, a sort of
modified quantum mechanics, in which wave function collapse propagates at the
speed of light \citep{akcll}.  Appleby notes the example of Newtonian gravity
himself, but states that ``in the framework of quantum mechanics the phenomena
of contextuality and non-locality are closely connected''
\citep[p.1]{appleby02}.  This is true, but it is not necessarily the case that
what is true in the framework of quantum mechanics is still true when we take
the point of view of the hidden variables --- and when assessing hidden
variable models, it is the hidden variables' point of view that is important.
\citet{appleby02} concludes, based on a GHZ-type example, that the CK models
display ``existential contextuality''.  It seems to us that, considered from
the proper hidden variable model theoretic rather than quantum theoretic
perspective, Appleby's argument simply demonstrates the non-locality of the CK
models --- which were, of course, explicitly presented as non-relativistic and
necessarily non-local.

Finally, some have objected to the MKC models on the grounds that elements of
the quantum formalism, for example the superposition principle \citep{cabello}
or the quantum logical relations between projectors
\citep{havliceketal,busch}, are not preserved. We note that this is of no
importance from the point of view of the hidden variables. The whole point is
that they have their own classical logical structure.

\section{Experimental tests of contextuality?}\label{experimentaltests}

Another issue that has arisen, both prior to and during the course of these
debates, is that of an experimental test of contextuality. Some experiments
have actually been performed. An examination of this issue, in particular of
what the experiments can really tell us, is of interest independently from the
MKC models and will improve our understanding of the Kochen-Specker theorem.
But the issue is also relevant for MKC models.  Indeed if it were possible to
rule out non-contextual theories via a decisive experimental test, this would
seem to contradict the claim that the CK models reproduce the predictions of
quantum mechanics to arbitrary precision and are non-contextual. In
Sec.~\ref{exptest} we argue that, quite independently of the issue of finite
precision, the idea of an experimental refutation of non-contextuality is
based on conceptual confusion, and that the experiments that have actually
been carried out are, as far as contextuality goes, not of major significance.
We examine in particular an experiment that has actually been performed,
\citet{hlzpg}, inspired by a proposal of \citet{szwz}, in turn based on a
scheme of \citet{cg}. (Another recent experiment is that of \citet{hlbbr}, which is
similar to a proposal of \citet{basuetal} - we do not discuss this in detail,
since the same arguments apply). In Sec.~\ref{finprecexp}, we argue that in
addition, the MKC finite precision loophole does apply, in the sense that any
experiment can be simulated by the CK models. Finally, in
Sec.~\ref{operationalapproach} we discuss the operational approach of
\citet{sbz} and \citet{larsson}.

\subsection{What can an experiment tell us about contextuality?}\label{exptest}

We begin by discussing the possibility of an experimental test of
contextuality in the absence of finite precision considerations. It is easiest
to do this with a particular example in mind, so we make particular reference
to the scheme which Simon et~al. (SZWZ) proposed and which inspired
the experiments reported by Huang et~al. (HLZPG).  
Consider a $4$-dimensional Hilbert
space, which we can think of as representing two $2$-dimensional subsystems.
The two subsystems are associated with the path and polarisation degrees of
freedom of a single photon. Define the subsystem observables $Z_1, X_1, Z_2,
X_2$, where subscript $1$ indicates the path degree of freedom and subscript
$2$ the polarisation degree of freedom.  Suppose that $\hat{Z}_i=\sigma_{zi}$
and $\hat{X}_i=\sigma_{xi}$, where $\sigma_{zi}$ and $\sigma_{xi}$ are Pauli
operators acting on subsystem $i$. Each of these observables can take the
values $+1,-1$. In an NCHV interpretation, a hidden state must assign a value
to each of these observables that would simply be revealed on measurement.
This in turn defines a colouring of the corresponding set of operators,
$V(\hat{Z}_1), V(\hat{X}_1), V(\hat{Z}_2), V(\hat{X}_2)$.

One can also consider observables that are products of these observables, for
example, $Z_1X_2$. Product observables also take the values $+1,-1$, and from
the KS criteria we have:
\begin{eqnarray}\label{productobservablecondition}
V(\hat{Z}_1\hat{Z}_2)&=&V(\hat{Z}_1)V(\hat{Z}_2)\nonumber\\
V(\hat{Z}_1\hat{X}_2)&=&V(\hat{Z}_1)V(\hat{X}_2)\nonumber\\
V(\hat{X}_1\hat{Z}_2)&=&V(\hat{X}_1)V(\hat{Z}_2)\nonumber\\
V(\hat{X}_1\hat{X}_2)&=&V(\hat{X}_1)V(\hat{X}_2)\, .
\end{eqnarray}
Finally, the contradiction arises on consideration of the quantum state
\begin{eqnarray*}
\ket{\phi_{+}} &=& \frac{1}{\sqrt{2}}(\ket{+z}\ket{+z}+\ket{-z}\ket{-z})\\
&=& \frac{1}{\sqrt{2}}(\ket{+x}\ket{+x}+\ket{-x}\ket{-x}),
\end{eqnarray*}
where $\ket{+z}$ is an eigenstate of $\hat{Z}_i$ with eigenvalue $+1$, and so
on. This state has the property that measurement of the product $Z_1Z_2$
always returns 1, as does measurement of $X_1X_2$. If
$V(\hat{Z}_1)=V(\hat{Z}_2)$, $V(\hat{X}_1)=V(\hat{X}_2)$, and
Eqs.~(\ref{productobservablecondition}) are satisfied, then it follows
logically that $V(\hat{Z}_1\hat{X}_2)=V(\hat{X}_1\hat{Z}_2)$. Yet in quantum
mechanics, one can measure $Z_1X_2$ and $X_1Z_2$ simultaneously, and if the
state is $\ket{\phi_{+}}$, then one will get opposite results with certainty.
Hence we have a contradiction.\footnote{This argument differs from
  standard Kochen-Specker-style proofs (and from Cabello and
  Garc\'{i}a-Alcaine's argument) in that the predictions from a
  particular quantum state are used to obtain a contradiction.}  

In principle, a
laboratory implementation could use a network of beam splitters, polarising
beam splitters and half-wave plates in order to prepare a single photon in the
state $\ket{\phi_+}$ and perform each of the joint measurements $$(Z_1,Z_2),
(Z_1,X_2), (X_1,Z_2), (X_1,X_2), (Z_1X_2,X_1Z_2) \, .$$

In the experiment of
HLZPG, only the $(X_1,X_2)$ and $(Z_1X_2,X_1Z_2)$ measurements were actually
performed, with the outcome of a potential $(Z_1,Z_2)$ measurement being
assumed from the method of state preparation.  
Though a detailed critique of HLZPG's experiment is beyond our scope here,
we should note that it deviates in various ways from the ideal version
proposed by SZWZ, and add that we find their discussion hard to
follow at various points: for example, they appear to interpret one
of their settings (their setup 2) as performing a simultaneous 
measurement of $X_1$, $Z_2$ and $Z_1 X_2$.

What, in any case, {\it could} an experiment along the lines suggested by
SZWZ show?  In each of
\citet{cg}, \citet{szwz}, and \cite{hlzpg}, the work is motivated via an
analogy with Bell's theorem. Bell's theorem tells us that locally causal
theories are incompatible with quantum mechanics, according to Bell's precise
definition \citep{belllc} of ``locally causal''.  The associated experimental
tests have strongly confirmed quantum mechanics. Then it is claimed, for
example, that
\begin{quotation}
  ``The Kochen-Specker theorem states that {\it non-contextual} theories are
  incompatible with quantum mechanics.'' \citep[p.1783]{szwz}
\end{quotation}
If one takes this at face value, it seems easy to accept that a
Kochen-Specker experiment to test non-contextuality would be of similar
interest and fundamental importance to a Bell experiment that tests local
causality.

However, there is a key point, not noted by these authors, where the analogy
breaks down.  A Bell experiment allows us to test the predictions of quantum
mechanics against those of locally causal theories because a definition of all
the terms used in a derivation of Bell's theorem (in particular the term
``locally causal'' itself) can be given that is \emph{theory-independent}. Yet
in the Kochen-Specker scheme above, the observables have not been defined in a
manner that is theory-independent, but have instead been defined with respect
to the quantum mechanical operators.  When a simultaneous measurement of
$Z_1X_2$ and $X_1Z_2$ is performed, the experimental setup as a whole looks
different from that employed in a simultaneous measurement of, say, $X_1$ and
$X_2$. 

For example, HLZPG describe two experimental setups: to get from one to
the other one needs to rotate the two half-wave plates they call HWP1 and HWP2.  
What gives us licence to claim that one of these setups really measures two 
observables, of which one is the product of $Z_1$ and
$X_2$ and the other is the product of $X_1$ and $Z_2$?  The answer is: our
conventional physical understanding of the experiment, \emph{as informed by
  the quantum formalism}. HLZPG need to assume that the effects of devices
such as beam splitters and half-wave plates are well described by the Hilbert
space formalism. That they do this implicitly is evident in remarks such as
``the interference on a BS [beam splitter] performs a Hadamard transformation
of the path qubit'' \citep[p.2]{hlzpg}. But there is no reason to assume that
such statements will be true (or even meaningful) in a theory that is not
quantum mechanics. Thus there is no theory-independent means of knowing that
we really are doing a simultaneous measurement of the product of $Z_1$ and
$X_2$, and the product of $X_1$ and $Z_2$. But this is crucial if we are to
conclude unequivocally that contextuality is being exhibited.
Similar comments apply to Hasegawa et al.'s experiment\citep{hlbbr}: their spin rotator and
phase shifter need to be adjusted to alter their parameters $\alpha$
and $\chi$, and they naturally need to rely on the standard quantum
formalism in order to interpret the experiment as carrying out 
measurements of particular projections onto the path and spin
degrees of freedom.    

Of course, the mathematical arguments given by these various 
authors are valid, and offer yet further
proofs that there are no NCHV interpretations of the quantum mechanical
formalism.  And clearly the experiments confirm some predictions of 
quantum theory. 
However, Cabello and Garc\'{i}a-Alcaine's claim 
that this type of experiment can show that
\begin{quotation}
``NCHV theories, \emph{without} any call to the formal structure of QM, make
conflicting predictions with those of QM'' \citep[p.1797, their emphasis]{cg},
\end{quotation}
which is echoed by HLZPG, is simply not correct.

These remarks apply quite generally to any proposed test of
contextuality that involves measuring product observables. 
Without using locality arguments, there is no way to 
guarantee that a given measurement is of an
observable that is precisely in product form, nor that two different 
measurements involve products of the same operator.  If such an experiment is performed,
and results consistent with quantum mechanics obtained, what can we conclude?
We have essentially three choices. First, accept the basic quantum formalism
and accept also that any underlying hidden variable theory assigning values to
Hermitian operators must be contextual. Second, look for loopholes in our
interpretation of the experimental results. Or third, reject the Hilbert space
structure and look for an entirely different theory of the experiment that is
non-contextual in its own terms.

The second move is exploited by the MKC models. The
third move will always be logically possible if non-contextuality is defined
(as it often is in the literature) as simply requiring that the value
obtained on measuring a given observable does not depend on which other
observables are measured at the same time. No mention of Hermitian operators
is given in this definition, so it has the appearance of being theory
independent. But it is not all that useful. It allows a non-contextual theory
of any experiment to be cooked up in a trivial manner, simply by redefining
what counts as an observable
--- for instance, by taking an observable to correspond to the full
projective decomposition of the identity defining any given measurement,
rather than to a single projection \citep{vf}.

Note that if a Bell experiment is performed, and the quantum predictions
verified, then we have analogues of the first two options above: we can
reject local causality, or we can look for loopholes in the experiment. Both
options have been much explored. But the analogy breaks down when we consider
the third option above, because the Hilbert space structure was not used
either in the derivation of Bell's theorem or in the interpretation of the
experiment.\footnote{
Of course, even local causality cannot be defined with {\it no}
assumptions about an underlying theory. It requires the notion of a
background space-time with a causal structure.  Bell's discussion of
the implications of local causality for Bell experiments also
implicitly requires that the notion of an experimental outcome has
its conventional meaning.  

It is worth noting, incidentally, that this last point 
leaves room for arguing that an 
Everettian interpretation of 
quantum theory might be defined so as to be locally causal.
We will not pursue this here, since the larger questions of 
whether a coherent Everettian interpretation exists, and if 
so on what assumptions, are beyond our present scope.} 
It also breaks down when we consider the outcomes of exploring 
the second option: finite experimental precision poses no fundamental
difficulty in the analysis of Bell experiments, but turns out to be an
unstoppable loophole in Kochen-Specker experiments. 

Granted then, that this type of experiment cannot be of decisive significance,
can it have \emph{any} significance?   Can it be interpreted 
as a test between quantum mechanics and a
different kind of theory? If it can, then it must be as a test between quantum
mechanics and non-contextual theories of a rather restricted kind. Such an
experiment, for example, could serve as a test between quantum mechanics and a
non-contextual theory that accepts some part of Hilbert space structure
(including the operators for path and polarisation degrees of freedom,and the
action of devices such as beam splitters), but rejects the KS criteria.
Logically, this would be a valid experiment. However, in order to motivate it,
one would need to devise an interesting and plausible alternative to quantum
theory which retains the features just mentioned but violates
(\ref{kscriteria}).  Considering such alternatives is beyond our scope here;
we only wish to note that the class of such alternatives is not nearly as
general and natural as the class of locally causal theories. So far as the
project of verifying the contextuality of Nature (as opposed to the
contextuality of hidden variable interpretations of the standard quantum
formalism) is concerned, the question is of rather limited relevance and
interest.

In conclusion, experiments along the lines of those of \citet{cg},
\citet{szwz} and \citet{hlzpg}, do not and cannot decisively distinguish
between contextuality and non-contextuality in Nature. If the quantum
formalism of states and operators (and the assignments of states and operators
to particular experimental devices) is not assumed, then the experiments tell
us little. On the other hand, if the standard quantum formalism is assumed,
then we know already from the Kochen-Specker theorem, before we carry 
out any experiments, that there is no
way of assigning values non-contextually to the set of all Hermitian
operators.  Mermin's comment that
\begin{quotation}
``the whole notion of an
experimental test of [the Kochen-Specker theorem] misses the
point'' \citep[Mermin, quoted in][]{cg} 
\end{quotation}
still seems to us to apply.

\subsection{Experiments and finite precision}\label{finprecexp}

In addition to the considerations of the last section, it is of course the
case that a CK model can simulate any quantum experiment, and this includes
so-called tests of the Kochen-Specker theorem. We shall leave it to the reader
to examine in detail how a CK model will work when applied to any specific
experimental setup. Obviously the fact that beam splitters, half wave plates
and so on, will be constantly shifting in alignment by minute amounts will
lead to finite precision in the case of the HLZPG experiment. This means that
each time a photon passes through the apparatus, the actual observables
measured will be slightly different. The CK models then show us that even if
it \emph{is} assumed that the operation of each experimental device is well
described by the Hilbert space formalism, a non-contextual, classical
simulation of the experiment is possible.\footnote{At the end of their paper,
  HLZPG make passing reference to the problem of finite precision, mentioning
  the work of \citet{sbz} and \citet{cabellopra}. The former we discuss below,
  here noting only that it is not relevant to HLZPG's experiment, since they
  do not actually apply the result, nor can it be applied to their data. The
  latter we have already mentioned in Sec.~\ref{classicality}, noting that the
  faithful measurement condition must be assumed, and that this is not
  necessary for classicality.}

We make a brief remark about the experiment of \citet{hlbbr}, and the proposal
of \citet{basuetal}. In both cases, an inequality is derived, formally
identical to the Clauser-Horne-Shimony-Holt inequality \citep{chsh}, that
concerns the spin and path degrees of freedom of a single neutron. It may seem
as if this evades the finite precision loophole, since the inequality is
violated by an irreducibly finite amount. The derivation of the inequality,
however, assumes that all measurements performed are strictly of the form
$A\otimes I$, in the case of a path degree of freedom, or $I\otimes B$, in the
case of a spin degree of freedom. A CK model, on the other hand, assumes that
the actual operators measured are not in fact precisely separable, even in
experiments which are designed to measure separate commuting observables.  
When arguments based on locality and space-like separation are forbidden
--- as they are here, since the question is whether quantum contextuality
can be demonstrated separately from quantum non-locality --- this is not 
physically implausible.  Beam splitters generally have a slight polarising effect, 
for example.  More generally, adjusting any piece of the experimental apparatus 
slightly influences all the others.  

\subsection{Defining observables operationally}\label{operationalapproach}

One may try to avoid the above arguments by framing a definition of
contextuality that is genuinely independent of Hilbert space structure. This
could be done by giving a completely operational definition of ``observable''
and hence of ``contextuality''. This may seem to have the additional advantage
of avoiding the issue of finite precision, since operational definitions do
not assume infinite precision in the first place. The operational approach is
hinted at in \citet{mermin99} and worked out explicitly by Simon, Bruckner and
Zeilinger (SBZ) and Larsson \citep{sbz,larsson}. The work of both SBZ and
Larsson is motivated by the issue of finite precision and is presented as a
riposte to MKC. SBZ, for example, describe their work as showing ``how to
derive hidden-variable theorems that apply to real experiments, so that
non-contextual hidden variables can indeed be experimentally disproved.''
This seems to contradict directly the claims of CK, in particular, who say
that the CK models are non-contextual and reproduce correctly the quantum
predictions for any finite precision experiment. We shall see, however, that
there is really no tension here. The apparent contradiction rests on different
uses of the word ``contextual''. Further, we shall argue that the work of SBZ
and Larsson, while interesting, does not have the significance they claim. For
definiteness, we discuss the work of SBZ, although Larsson's is very similar.

SBZ consider a black box with three knobs, each of which has a finite number
of different settings. After setting the knobs, an observer presses a ``go''
button. He then receives an outcome for each knob, which is either a $1$ or a
$0$.  As an example of such a box, we can consider one that contains within
it a quantum experiment in which the spin squared of a spin-$1$ particle is
measured in three different directions. The directions are determined to some
degree of accuracy by the settings of the knobs. However, it will not be the
case that a given knob setting corresponds to a measurement of spin squared
in precisely the same direction every time the box is used. There will be
experimental inaccuracies. In general, we may imagine that there are some
hidden variables associated with the measuring apparatus, as well as the
quantum system, which determine exactly what measurement is being performed.
From the point of view of our observer outside the black box, however, none
of this matters. All he has access to are the three knobs and the outcomes.
SBZ propose that the observer should simply, by fiat, define observables
operationally, with each observable corresponding to a different setting of
one of the knobs. He can always be sure which observable he is measuring,
according to this operational definition, even though he cannot be sure which
observable is actually being measured according to quantum theory.

Not knowing what is happening inside the box, our outside observer can try to
formulate a model theory. In a deterministic model theory, the entire inside
of the box can be described by some hidden state that predicts what the three
outcomes will be for each possible joint setting of the knobs. The model is
non-contextual if, for each hidden state, the outcome obtained for each knob
depends only on its setting, and not on the settings of the other two knobs.
On running the box repeatedly, the observer can build up outcome statistics
for each possible joint knob setting. If no non-contextual model of the
workings of the box that reproduces these statistics exists, for any
deterministic model theory, then, SBZ
propose, we should say that the box is ``contextual''.  

Let us consider the implications of this definition applied to projective
measurements on a $3$-dimensional space.   Take a set 
of $3$-dimensional vectors that is KS-uncolourable, in the
sense that it is impossible to give each vector a $0$ or a $1$ such that each
orthogonal triad consists of one $1$ and two $0$s.  The set of vectors can be
written, for example
\[
\{\{\vec{n}_1,\vec{n}_2,\vec{n}_3\},\{\vec{n}_1,\vec{n}_4,\vec{n}_5\},\ldots\}.
\]
For the set to be KS-uncolourable, it must be the case that some vectors
appear in more than one triad. Suppose that these triads are taken to
indicate possible triads of knob settings. Suppose that the experiment is run
many times, and it is found that whenever one of these triads is measured,
the outcomes consist of one $1$ and two $0$s. Then we can conclude, from the
fact that the set of vectors is KS-uncolourable, that the box is
``contextual'' according to SBZ's definition --- a property we
refer to hereafter as SBZ-contextual. 

This, though, is too much of an idealisation. In a real experiment there will
be noise, which will sometimes cause non-standard results, for example two
$1$s and a $0$. The core of SBZ's paper is a proof of the following result.
Imagine that the box is run many times, with knob settings corresponding to
orthogonal triads, and that the outcomes are one $1$ and two $0$s in a
fraction $1-\epsilon$ of cases. Then, the box must be SBZ-contextual if
$\epsilon<1/N$, where $N$ is the number of orthogonal triads appearing in the
set. If the box is in fact a quantum experiment in which the spin squared of
a spin-$1$ particle is measured in different directions, then increasing the
accuracy of the experiment will be able to reduce $\epsilon$ below $1/N$. The
observer will be able to conclude that the experiment is SBZ-contextual.

We wish to make several related remarks concerning this result. The first
thing is to clarify the implication for MKC models. A box with a quantum spin
experiment inside is certainly simulable by a CK model, since the models
are explicitly constructed to reproduce all the predictions of quantum
mechanics for finite precision measurements. How will the simulation work? On
each run, the knob settings determine approximately which measurement is
performed, but exactly which is determined randomly, or by apparatus hidden
variables. The exact measurement corresponds to some Hermitian operator in
the CK KS-colourable set. The outcome is determined by a hidden state that
assigns a definite value to each operator in the KS-colourable set in a
non-contextual manner. Hence if observables are defined by operators, it is
true that the value obtained on measuring a given observable does not depend
on which other observables are measured at the same time and in this sense,
the CK model is non-contextual. The fact that the black box is SBZ-contextual 
tells us that the settings of all three knobs together,
along with the apparatus hidden state, are needed to determine the Hermitian
operators that are in fact being measured.  In a way, of course, it couldn't
be any different, since one cannot expect an algorithm that chooses three vectors 
independently generally to produce an orthogonal triad. 
The SBZ-contextuality of the black box
tells us in addition that for at least some apparatus hidden states, whether
the measurement corresponds to a triad for which knob $i$ gets outcome $0$ or
a triad for which knob $i$ gets outcome $1$ depends on the settings of knobs
$j$ and $k$.

This should be enough to show that there is no formal contradiction between
the CK and the SBZ results. Some may argue, however, that from a physical
point of view, the operational definition of SBZ-contextuality is the only
interesting one, and that the CK models, therefore, are not non-contextual
in any interesting sense --- or at least that the operational definition is
an interesting one, and the CK models are not non-contextual in this sense.
We wish to counter such arguments with some cautionary remarks concerning
these black boxes.

First, SBZ, as did the authors of the experiments discussed in
Sec.~\ref{experimentaltests} above, motivate their work via an analogy with
Bell's theorem. The disanalogy we mentioned in Sec.~\ref{experimentaltests}
has disappeared now that observables are defined operationally. However,
there is another important disanalogy. This is that there is nothing
specifically non-classical about a black box that is behaving SBZ-contextually.
One could easily construct such a
box out of cog-wheels and springs.  Thus with no knowledge of or assumptions
about the internal workings of the box, one could not use it to distinguish
classical from quantum behaviour.  

This should be contrasted rigorously with
the case of a black box which behaves non-locally in the sense
of producing Bell correlations.  (Of course, to define this sense of non-locality,
we assume that special relativity is approximately correct and an approximately
Minkowski causal structure is given.)  Then, if we have a black box, large enough to allow 
space-like separated outputs to be identified, which behaves non-locally, 
we know that we are in a quantum, and not a 
classical, universe. Such a box can even be used for information theoretic
tasks that cannot be accomplished classically \citep[e.g.,][]{bcvd}. Given
a black box that is SBZ-contextual, we have no such 
guarantees. This seems to us to cast doubt on the use or 
significance of a purely operational definition of
contextuality, as opposed to a theory-relative one.

Second, the fact of the matter is that any realistic experiment, whether carried
out in a classical or a quantum universe, will necessarily 
exhibit SBZ-contextuality to some (possibly tiny) degree.  Not
only that, the {\it outcome probabilities} for any given SBZ-observable will depend
(at least slightly) on the context of the other knob settings. On moving one
knob, for example, its gravitational field will be changed, and this will
affect the behaviour of the whole apparatus.  This is not a consequence of
quantum theory.  It would be true of an experiment in which a classical
measuring apparatus measures classical observables on a classical system. Yet
we would not infer from this SBZ-contextuality of the outcomes that classical
physics is (at least slightly) contextual. We do not take SBZ and Larsson to
be advocating otherwise: all sides in the Kochen-Specker debate agree that
classical physics is, paradigmatically, non-contextual. Rather, we take the
fact that the opposite conclusion follows from SBZ's and Larsson's definitions
to indicate that the definition of SBZ-contextuality is inherently flawed.
Similarly, we take the fact that SBZ's definition of an observable can in 
principle 
empirically be shown to be context-dependent --- since the outcome
probabilities depend at least slightly on knob settings that are meant to
correspond to independent observables --- to be a fatal flaw in that
definition. An SBZ-observable turns out, under scrutiny, to be a rather
complicated construct, with quite different properties from its quantum
namesake.  A less freighted 
name --- ``dial setting'', for instance --- would
make clearer the obstacles which SBZ would need to surmount in order even to
begin a properly founded discussion of finite precision experimental tests
of contextuality.

This last point really needs no reinforcement, but it can be reinforced.
Consider again the black box that in fact contains a quantum experiment in
which the spin squared of a spin-$1$ particle is measured in different
directions. The idea was to run the box repeatedly with certain combinations
of knob settings that correspond to the orthogonal triads in a
KS-uncolourable set of vectors. However, assuming that they can be moved
independently, there is nothing to stop us from setting the knobs in any
combination of settings, in particular, in combinations that correspond to
triads of non-orthogonal vectors from the KS-uncolourable set. What would
happen in this case? The quantum experiment inside the box cannot be
effecting a simultaneous measurement of the spin squared in three directions
approximating the knob settings, because these spin squared observables will
not be co-measurable. Perhaps the box measures spin squared in three
orthogonal directions, at least one of which is not close to the
corresponding knob setting. Or perhaps the box does some kind of positive
operator valued measurement. In either case, it seems that for most quantum
experiments, from the observer's point of view, the outcomes will inevitably
be contextual even at the level of the quantum probabilities, and even if we
unrealistically neglect the classical perturbations produced on the apparatus
by altering any of the knobs.  Given that the
box is behaving in an overtly contextual manner even at the level of
probabilities, one is then again led to ask: why should we be interested in whether the box can be
described in a non-contextual fashion in the special case that we carefully
restrict our knob settings so that they always correspond to orthogonal
triads in the KS-uncolourable set?

Taking these points on board, careful operationalists might try to refine 
their position by speaking, not of a distinction between SBZ-contextuality and
SBZ-non-contextuality, but instead of degrees of SBZ-contextuality. It could be
argued that, although classical mechanics is indeed SBZ-contextual, the
perturbations that imply SBZ-contextualities in outcome probabilities will
generally be very small, and the outcome probability SBZ-contextualities
correspondingly hard to detect: indeed, in principle, with
sufficient care, the perturbations can be made as small as desired.  In
contrast, SBZ and Larsson's results might be interpreted as 
implying that quantum experiments display an irreducible finite degree of
SBZ-contextuality.  The difficulty with this line of argument is
that, as the CK models illustrate, it is {\it not} always true in 
classical mechanics that small
perturbations induce (only) correspondingly subtle effects. 
Operationalists would need to frame a
definition which separates classical mechanics not only from the CK models
(defined by infinite subsets of projective or positive operator valued
decompositions) but also from infinitely many of their finite sub-models (defined,
as above, by finite subsets which reproduce quantum theory to some fixed 
finite precision $\epsilon_r$), in order 
to maintain both that classical mechanics is at least approximately or effectively
SBZ-non-contextual and that all finite precision 
approximations to quantum theory defined by CK models that are
precise to within $\epsilon$, for some $\epsilon > 0$, are definitively SBZ-contextual.  
This cannot be done: as we have already noted, in principle the finite sub-models give 
a prescription for building real classical devices with finitely many
degrees of freedom, and these devices are, of course, described by 
classical mechanics. 

In summary, even black box operational definitions do not allow unambiguous
experimental discrimination between contextual and non-contextual theories,
and thus present no challenge to CK's assertion that non-contextual theories
can account for current physics.  SBZ's operational definition of
contextuality does give us a clear, theory-independent notion of
something, but it is not contextuality in any sense consistent
with standard usage.  In particular, the notion defined is not able to separate the properties of
quantum theory and classical mechanics, and so is not of fundamental
relevance to the debate over finite precision and the KS theorem.  Attractive
though it would be to devise a sensible theory-independent definition of
(non-)contextuality, we do not believe it is possible.  We see no
fundamentally satisfactory alternative to restricting ourselves to talking of
\emph{theories} as being non-contextual or contextual, and using
theory-relative definitions of these terms.

\section{A Closing Comment} 

We would like to emphasise that neither the preceding discussion
nor earlier contributions to this debate (\citet{akprl,ck}) are or 
were intended to cast doubt on the 
essential importance and interest of the Kochen-Specker theorem.  
As we have stressed throughout, our interest in examining
the logical possibility of non-contextual hidden variables
simulating quantum mechanics is simply that it {\it is} a 
logical --- if scientifically highly implausible ---
possibility, which demonstrates interesting limitations on what 
we can rigorously infer about fundamental physics.  

\vskip 10pt 
\leftline{\bf Acknowledgements}

We are profoundly indebted to Rob Clifton for many lively and
stimulating discussions and much encouragement. 
We also warmly thank Marcus Appleby, Jeremy Butterfield, Ad\'an Cabello, Chris Fuchs,
Sheldon Goldstein, Lucien Hardy, Jan-{\AA}ke Larsson, David Mermin, David Meyer,
Asher Peres, Christoph Simon and Jos Uffink for very helpful discussions and criticisms. 
This research was partially funded by the projects PROSECCO (IST-2001-39227) and 
RESQ (IST-2001-37559) of the IST-FET programme of the EC, the Cambridge-MIT
Institute and an HP Bursary.


\end{document}